\documentclass[twocolumn]{article}
\usepackage{graphicx} 
\usepackage{amsmath} 
\usepackage{authblk} 
\usepackage{blindtext}
\title{\textbf{Radio emission in the Jovian magnetosphere driven by the quasi-linear diffusion}}
\author[1]{A. Chkhaidze}
\author[2]{E. Kalichava}
\affil[1,2]{Unaffiliated}

\date{} 
\begin{document}

\twocolumn[
\maketitle

{\Large\textbf{Abstract}}

Electrons traveling along magnetic field lines from Jupiter to Io, driven by quasi-linear diffusion (QLD), emit synchrotron radiation. 
By using the small angle approximation, the kinematic equation for the particle distribution gives us the mean pitch angle value.
We described how these ultrarelativistic electrons emit radiation in the GHz range when subjected to external forces.
Despite varying Lorentz factors, we showed that the total QLD-related radiation power is in multiples of $10^{17} ergs/s$.
However, when this radiation reaches the Earth’s surface, its intensity falls within the range of around $10^{-7} erg/(s \cdot cm^{2})$.
\\
]

\section{Introduction}
The paper is organized as follows: in Sec. 2 we summarize the general theory of resonant acceleration. In Sec. 3, we apply the mechanism to Jupiter-Io system, derive and discuss the principal results. In Sec. 4, we put our results in perspective.
\section{Theoretical model}
Firstly, we need show table of physical parameters\textbf{(Table 1)}, which we will be using in all of our calculations, figures and tables.
\begin{table}[htbp]
    \centering
    \caption{Physical parameters}
    \label{tab:math-table-1}
    \begin{tabular}{cc}
        \hline
        $Name \, of \, parameter$ & $Value \, of \, parameter$ \\
        \hline
         speed of light $(c)$ & $2.9972*10^{10}(cm/s)$   
            \\
            electron mass $(m)$    & $9.1094 * 10^{-28}(gram)$
            \\
            electron charge $(e)$    & $4.8032 *10^{-10}(franklin)$      \\
            maximum magnetic field $(B_0)$    & $19.6631 (gauss)$
            \\
            curvature radius $(\rho)$ & $2.1297*10^{10}(cm)$ \\
            average radius of Jupiter $(R_j)$ & $6.9911*10^9(cm)$\\
            Jupiter's magnetic moment $(\vec{M})$ & $1.55*10^{30}(gauss*cm^3)$\\
            plasma Lorentz Factor $(\gamma_p)$ & $1$ \\
            Boltzman constan $(k_{boltzman})$ &  $1.3806*10^{-23}\bigg(\dfrac{J}{k}\bigg)$\\
            Distance from Earth to Jupiter $(R_e)$ & $8.5872*10^{13}(cm)$\\
            Effective radius $(R_{eff})$ & $25000 (cm)$\\
            System noise $(T_{system})$ & $20 (k)$\\
        \hline
    \end{tabular}
\end{table}

Now we start by modeling Jupiter's magnetic field. Jupiter's magnetic field is dipolar (\cite{connerney},\cite{warwick}) and to approximate the magnetic field strength we will use the following formula:
 \newline
 \begin{equation}
       \vec{B} = \alpha \bigg(\dfrac
{3\vec{r}(\vec{M}\cdot \vec{r}) - \vec{M}
r^2}{r^5}
\bigg)
 \end{equation}
 $\vec{r}$ is radius vector measured from the center of Jupiter and $\vec{M} = 1.55 * 10^{30} G\cdot cm^3$ is the Jupiter's magnetic moment as seen in \cite{russell}. This formula is normalised so that it coincides with empirical values of the induction, $\alpha\approx 2.3$ is the normalisation coefficient. We can use the following parametrization using distance $s$ along the trajectory, starting from the center of the planet. 
 \newline
    \begin{equation}
   \begin{array}{l}
     \theta = \dfrac{s}{2\rho} \\
      r = 2\rho \sin\bigg(\dfrac{s}{\rho}\bigg)
   \end{array}
   \end{equation}

\begin{figure}[htbp]
    \centering
    \includegraphics[width=0.5\textwidth]{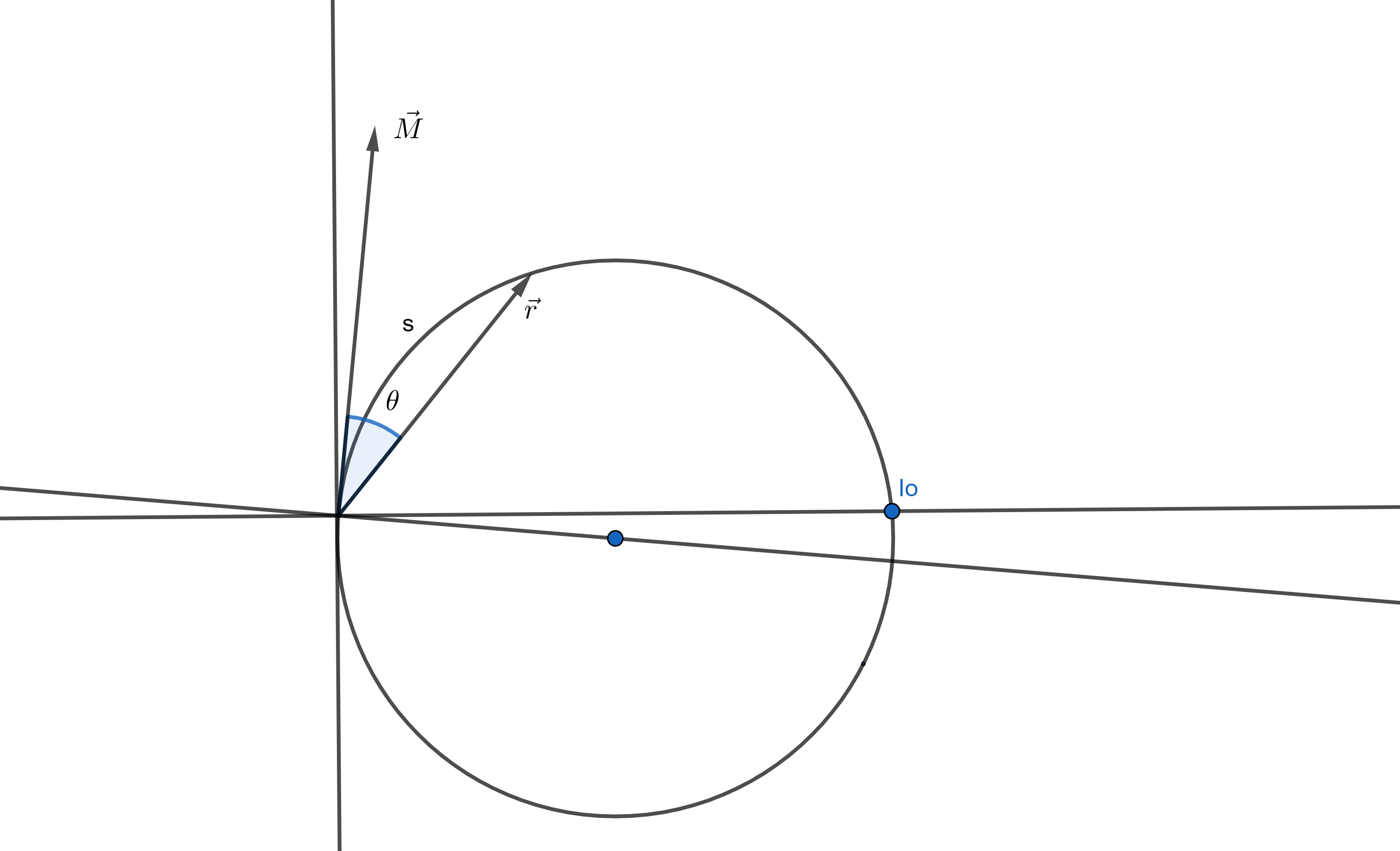} 
    \caption{Geometry of electron trajectory starting from Jupiter's center to Io and back to Jupiter. $\vec{M}$ is magnetic moment of Jupiter. $\vec{r}$ is radius vector measured from Jupiter's center. $s$ is distance along trajectory and $\theta$ is angle between $\vec{r}$ and $\vec{M}$}
    \label{fig:example-1}
\end{figure}
 Also we need to model the empirical data of corotating electron's concentration along IFT(Io-Flux Tube) as in \cite{bolton}.
 \newline 
 \begin{equation} 
    n_p(s) = 1.1 + 8 \cdot 10^{-4}e^{\bigg(2\bigg(\dfrac{s-1.02R_j}{R_j}\bigg)\bigg)}+ 
\end{equation}
\begin{equation*}
    +10^{5}e^{\bigg(-32\bigg(\dfrac{s-1.02R_j}{R_j}\bigg)\bigg)} cm^{-3}
\end{equation*}
Here,  $R_j$ is mean radius of Jupiter. The dependence of the number density of corotating electrons $n_p(cm^{-3})$ on parameter $s$ is depicted on \textbf{Fig.2} 

\begin{figure}[h]
    \centering
    \includegraphics[width=0.5\textwidth]{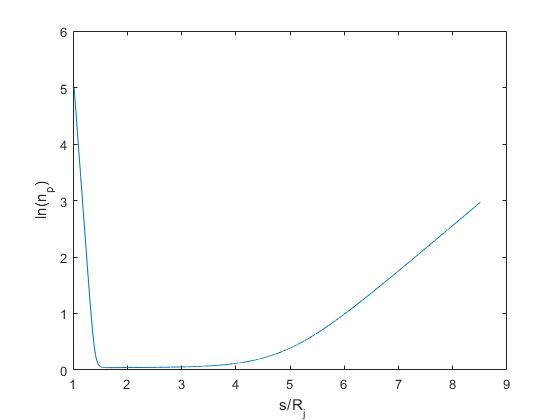} 
    \caption{Number density of corotating electrons $n_p$ is in $cm^{-3}$ . The distance along the trajectory is denoted by $s$ $(cm)$. This figure is in agreement with empirical data. It is clear that the electrons have highest density at the surface of Jupiter. Density quickly decreases as we leave Jupiter's Ionosphere. As we approach the Io torus the concentration increases significantly due to the presence of electrons originating from Io's volcanic activity. Parameters are taken from Table 1}
    \label{fig:example}
\end{figure}
\newpage
 As can be seen in \cite{osmanov} for QLD(Quasi-Linear Diffusion) to be present a certain criteria must be met. Electrons must be in a frozen-in condition. 
 \begin{equation}
    \gamma m c^2 n\leq \dfrac{B^2}{8\pi}
 \end{equation}

 Condition (4) will restrict electron beam concentration. According to \cite{gogaberishvili} during the frozen-in condition plasma may be subjected to the anomalous Doppler effect. This induces unstable resonance cyclotron modes. 
 
 \begin{equation}
    \omega - k_{||}c - k_x u_x - \dfrac{\omega_B}{\gamma_b} = 0
\end{equation}
Where $k_x$  and $k_{||}$ are the wave vector components along the drift and magnetic field lines respectively. $\gamma_b$ is beam Lorentz factor. $u_x$ is curvature drift velocity. $\omega_B$ is cyclotron frequency. 
\begin{equation}
   \begin{array}{l}
      u_x\approx \dfrac{c^2 \gamma_b}{\rho \, \omega_B}\\
      \\
      \omega_B = \dfrac{eB}{mc}
   \end{array}
 \end{equation}
 Due to the resonance we may have QLD, which will preserve non-zero pitch angle \cite{gogaberishvili}. Hence synchrotron emission mechanism is conserved.
 As can be seen in \cite{osmanov} using small angle approximation the kinematic equation for the particle distribution function can be reduced to the following form:

  \begin{equation}
     \dfrac{\partial f}{\partial t} + \dfrac{1}{p_\bot}\dfrac{\partial}{\partial p_\bot}(p_\bot[F_\bot + G_\bot]f^0)= 
 \end{equation}
 \begin{equation*}
     =\dfrac{1}{p_\bot}\dfrac{\partial}{\partial p_\bot}\bigg(p_\bot D_{\bot,\bot}\dfrac{\partial f^0}{\partial p_\bot}\bigg)
 \end{equation*}
\begin{equation}
     \begin{array}{l}
          D_{\bot,\bot} = D \delta E_{k}^{2}  \\
          \newline\\
          |E_{K}|^2 = \dfrac{mc^3 n_b \gamma_b}{4\pi\nu} \\
          \newline\\
          \nu \approx \dfrac{\omega_{B}}{2\pi\delta\gamma_{b}}\\
          \newline\\
          \delta = \dfrac{\omega_{P}^{2}}{4\omega_{B}^{2}\gamma_{p}^{3}}\\
          \newline\\
          \omega_{p} = \sqrt{\dfrac{4\pi n_{p}^{2}}{m}}
     \end{array}
 \end{equation}
Where $D_{\bot,\bot}$ denotes the diffusion coefficient and $D = \dfrac{e^2}{8c}$ . Here we also used $E_{k}^{2}$, which is the wave energy density per unit wavelength. Assuming that the half of the beam's energy is converted to that of the waves, we get the formula mentioned above. $\omega_{p}$ is the Langmuir frequency of the electron-positron plasma component and $n_{p}$  is the corresponding number density. 
For small angle approximation force due to the synchrotron radiation may be written as: \cite{osmanov}:
\begin{equation}
 \begin{array}{l}
         F_{\bot} = - \alpha_{s}\dfrac{p_{\bot}}{p_{\parallel}}\left( 1 + \dfrac{p_{\bot}^2}{m^2c^{2}} \right)\\
         \newline\\
          F_{\parallel} = - \dfrac{\alpha_{s}}{m^{2}c^{2}}p_{\bot}^{2}\\
          \newline \\
          \alpha_{s} = \dfrac{2e^{2}\omega_{B}^{2}}{3c^{2}}
 \end{array}
 \end{equation}
 
Where $p_\bot$ and $p_{||}$ are transversal and longitudinal components of momentum of electron. It should be noted that adiabatic invariant $I = \dfrac{3cp_\bot^2}{2eB}$ \cite{lominadze} is conserved because magnetic field changes insignificantly during the cyclotron period. As a result of this, particles also undergo the force responsible for the conservation of the adiabatic invariant.
 \begin{equation}
     \begin{array}{l}
           G_{\bot} = - \dfrac{cp_{\bot}}{\rho} \\
            \newline \\
            G_{\parallel} = \dfrac{cp_{\bot}^{2}}{\rho p_{\parallel}}
     \end{array}
 \end{equation}

 Where $\rho$ is curvature radius of the trajectory. Also we assume that different currents of electrons have the same energy \cite{gogaberishvili}.

 \begin{equation}
      n_{p}\gamma_{p} = n_{b}\gamma_{b}
 \end{equation}

 We take into account that  $\dfrac{{F_\bot}}{{G_\bot}}$ has the form as seen in (12).
 
 \begin{equation}
     \dfrac{F_\bot}{G_\bot} \approx 1.3311 \cdot 10^{-8} \bigg(\dfrac{\rho}{2.1297\cdot 10^{10} cm}\bigg)\cdot
 \end{equation} 
 \begin{equation*}
      \cdot\bigg(\dfrac{B_0}{19.6631 Gauss}\bigg)^2 \bigg(\dfrac{40}{\gamma}\bigg)
 \end{equation*}
 If we input the values of curvature radius, magnetic field at the surface of Jupiter and $\gamma = {40,60,80,100}$. We get that $\dfrac{F_\bot}{G_\bot} << 1$. Hence $F_\bot$ can be disregarded in equation (7). Also the duration in which we observe the electrons ($\approx 10^0$ s) is substantially smaller than the period of rotation of Io ($\approx 10^5$ s) . Therefore we can consider the process to be quasi-static. Hence equation (7) reduces to the following form.
 \begin{equation}
    \dfrac{1}{p_{\bot}}\dfrac{\partial}{\partial p_{\bot}}\left( p_{\bot}\left\lbrack G_{\bot} \right\rbrack f^{0} \right) = \dfrac{1}{p_{\bot}}\dfrac{\partial}{\partial p_{\bot}}\left( p_{\bot}D_{\bot,\bot}\dfrac{\partial f^{0}}{\partial p_{\bot}} \right)
\end{equation}
 Solving (13) for the distribution function we get the following result:
 \begin{equation}
    f^{0}\left( p_{\bot} \right) = C\exp\left( - \left( \dfrac{p_{\bot}}{\sqrt{\dfrac{2\rho D_{{\bot,\bot}}}{c}}} \right)^{2} \right)
\end{equation}
 Using (14) we can calculate the mean pitch angle at the starting point of trajectory.
 \begin{equation}
      \overline{\psi_{0}} = \dfrac{\overline{p_{\bot_{0}}}}{p_{\parallel}} = \dfrac{\int_{0}^{\infty}{p_{\bot}f^{0}\left( p_{\bot} \right)p_{\bot}}}{p_{\parallel}\int_{0}^{\infty}{f^{0}\left( p_{\bot} \right)p_{\bot}}} = \dfrac{\sqrt{\dfrac{2\rho D_{\bot,\bot}}{c}}}{p_{\parallel}\sqrt{\pi}} = \dfrac{p_{\bot_{0}}}{p_{\parallel}}
 \end{equation}
  Using the fact that adiabatic invariant  is conserved, we can calculate how the pitch angle changes along the whole trajectory.  
    \begin{equation}
        p_\bot(s) = \sqrt{\dfrac{B(s)}{B_0}} p_{\bot 0}
    \end{equation}
    \begin{equation}
        \psi(s) = \sqrt{\dfrac{B(s)}{B_0}} \overline{\psi_{ 0}}
    \end{equation}
Using parameters in \textbf{Table 1} and formula (17) we get the values of the pitch angle along the trajectory \textbf{Fig 3.} The apparent decrease of pitch angle is to be expected from formula (17) since magnetic field strength decreases along the trajectory. Parameters are taken from Table 1. 
\begin{figure}[htbp]
    \centering
    \includegraphics[width=0.5\textwidth]{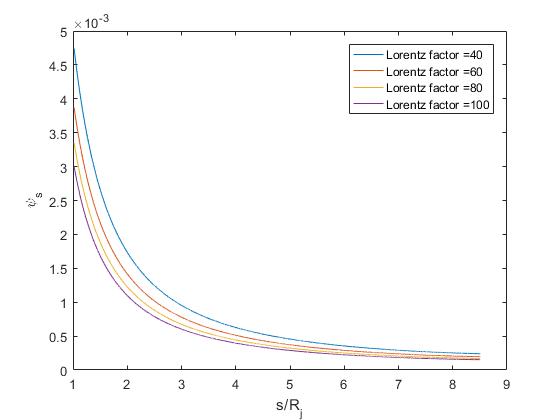} 
    \caption{$\psi$ is the pitch angle measured in $radians$. The distance along the trajectory is denoted by $s$ $(cm)$. The apparent decrease of pitch angle is to be expected from formula (17) since magnetic field strength decreases along the trajectory. Parameters are taken from Table 1}
    \label{fig:example}
\end{figure}
 We use synchrotron radiation formula\cite{rybicki} to calculate emitted power from single electron.
    \begin{equation}
        P(s)=\dfrac{2e^4\gamma^2B^2sin^2\psi}{3m^2c^3}
    \end{equation}
 Also synchrotron emission critical frequency has the form \cite{rybicki}:
 \begin{equation}
     \omega_c =\dfrac{3\gamma^2qBsin(\psi)}{2mc}
 \end{equation}
This gives us the following formula for photon frequency \cite{osmanov}:
\begin{equation}
    f_c \approx \dfrac{3\gamma^2qBsin(\psi)}{4\pi mc}  
\end{equation}
Taking this into account small angle approximation we get the following normalised formulas of photon frequency. 
\begin{equation}
    f_c \approx 6.2665 \cdot 10^8 \cdot\dfrac{\gamma^2}{1600} \cdot\dfrac{B}{19.6631 Gauss} \cdot\dfrac{\psi}{0.0047} Hz
\end{equation}

\section{Results}
Now we show the results of calculations of total power and total intensity on earth's surface for different values of Lorentz factor. Firstly, we show the expected emission frequencies from equation (20) in \textbf{Fig 4}. Parameters are taken from Table 1. As we can see from the figure frequency of emitted photons decrease since both $B$ and $\psi$ decrease along the trajectory. The observations of Jupiter’s radiation spectrum have confirmed the existence of photons with frequencies of several GHz \cite{bolton}. The data also shows the existence of photons in the lower frequency range starting from 5 KHz \cite{kurth}. Thus, our calculations can explain this type of radiation in this frequency range.

\begin{figure}[htbp]
    \centering
    \includegraphics[width=0.5\textwidth]{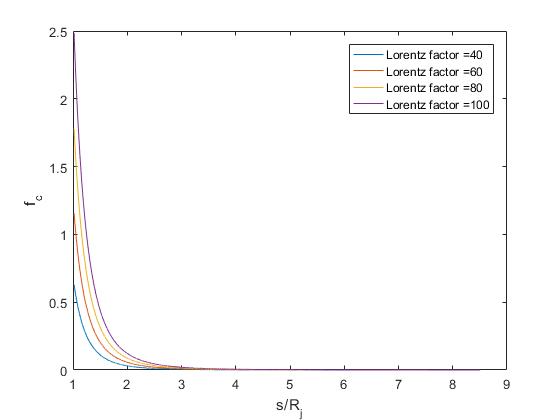} 
    \caption{We show the dependence of $f_c(GHz)$, which is the frequency of emitted photons, on $s(cm)$, which is the electron's distance along the trajectory. Parameters are taken from Table 1}
    \label{fig:example}
\end{figure}
 \textbf{Fig 5} shows how concentration of electrons in a beam $n_b(cm^{-3})$ is dependent on distance along the trajectory $s(cm)$. It is clear that we have a significant change of concentration which is due to restriction which we place on the beam due to frozen-in condition. Parameters are taken from Table 1. 
\begin{figure}[htbp]
    \centering
    \includegraphics[width=0.5\textwidth]{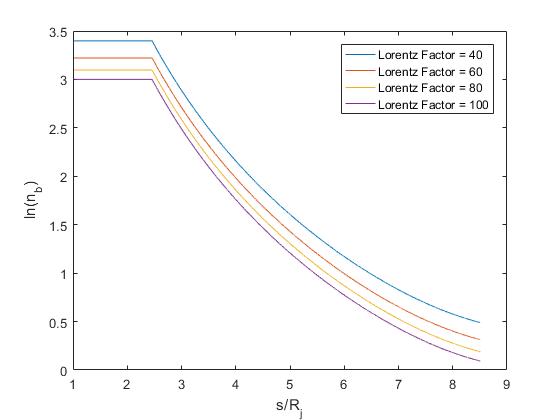} 
    \caption{$n_b$($cm^{-3}$) is concentration of electrons in a beam with specific Lorentz Factor , $s (cm)$ is distance along trajectory. As can be seen from the figure at approximately $s = 2.46 R_j$ we have a significant change of concentration which is due to restriction, which we place on the beam due to frozen-in condition. Parameters are taken from Table 1}
    \label{fig:example}
\end{figure}
\textbf{Fig 6} shows how power $P(ergs/s)$ emitted by a single electron from equation (18) changes along the trajectory. Parameters are taken from Table 1. 
\begin{figure}[htbp]
    \centering
    \includegraphics[width=0.5\textwidth]{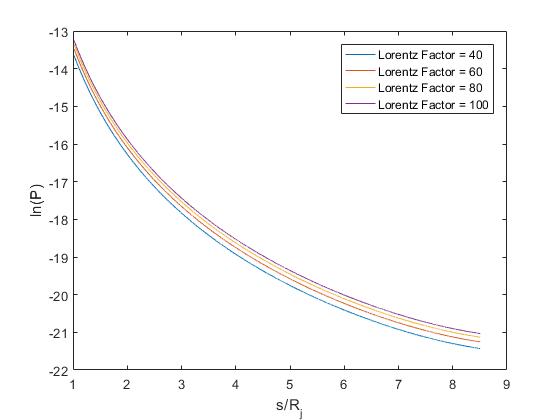} 
    \caption{$P(ergs/s)$ is power emitted per electron, $s (cm)$ is distance along trajectory. $P$ is decreasing which is to be expected since both $B$ and $\psi$ in equation (18) are decreasing. Parameters are taken from Table 1.}
    \label{fig:example}
\end{figure}

\begin{figure}[htbp]
    \centering
    \includegraphics[width=0.5\textwidth]{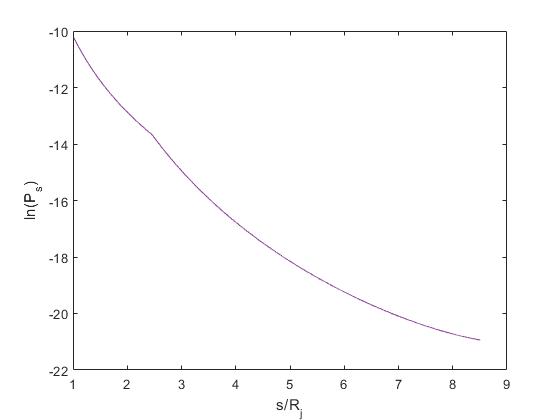} 
    \caption{$P_s (ergs/s)$ is power emitted by electrons in  $1 cm^3$ at given $s(cm)$ distance along the trajectory  At approximately $s= 2.4 R_j$ there is a change in a function behaviour. This  is a result of a sudden decrease of concentration $n_b$, which is caused by the restriction of frozen-in condition. Parameters are taken from Table 1.}
    \label{fig:example}
\end{figure}

\newpage
Afterwards, we calculate the power emitted by electrons in $1cm^3$ along the trajectory of motion, using the following relation: $P_s = P \cdot n_b$ \textbf{(Fig 7.)}
At approximately $s= 2.4 R_j$ there is a change in a function behaviour. This  is a result of a sudden decrease of concentration $n_b$, which is caused by the restriction of frozen-in condition (\textbf{Fig. 5}). Parameters are taken from Table 1.

The power emitted by ultrarelativistic electrons in definite volume can be given as:
\begin{equation}
    P_{total} = \int_{V}^{} P_{s}dV 
\end{equation}
As a source of radiation we will consider electrons in a torus segment with one end at $s = 1.02 R_j$ and another end at $s = 2.4 R_j$ with base that has radius  $R_j \cdot sin(10$\textdegree$)$. The total radiation power generated by this volume would be $P_{total1} = 2.3010 \cdot 10^{17} (ergs/s)$. If we consider torus segment ending at $s \approx 8.5R_j$(place where a center of Io torus or Io is located) with the same base radius, the numerical result for total power would be $P_{total2} = P_{total1} + 1.7143 \cdot 10^{14}(ergs/s) $. As we can see difference between $P_{total2}$ and $P_{total1}$ is insignificant compared to $P_{total1}$. Hence we will be discussing volume between $s=1.02R_j$ and $s = 2.4R_j$. Total power emitted by the volume of interest for different Lorentz factors are calculated using equation (22). Results are shown in \textbf{Table 2.} Parameters are taken from Table 1.
   \begin{table}[h]
      \caption[]{Total power for different Lorentz factor}
         \label{}
     $$ 
         \begin{array}{p{0.4\linewidth}l}
            \hline
            \noalign{\smallskip}
            Lorentz factor      &  Power(ergs/s)  \\
            \noalign{\smallskip}
            \hline
            \noalign{\smallskip}
            40 & 2.30096 * 10^{17}   
            \\
            60    & 3.45146 * 10^{17}
            \\
            80    & 4.60197 * 10^{17}      \\
            100   & 5.75248 * 10^{17}
            \\
            \noalign{\smallskip}
            \hline
         \end{array}
     $$ 
   \end{table}
   \newline
Since we have relativistic electrons, the radiation experiences beaming effect, meaning that very few photons will be emitted having pitch angle $\psi\geq$ $\dfrac{1}{\gamma}$. Hence, $R_e$ distance away from the source,  the power radiated  will be dispersed on surface with area equal to $\pi\bigg(\dfrac{R_e}{\gamma}\bigg)^2$. So the intensity can be written as equation (23).
\begin{equation}
    I = \dfrac{P}{\pi \bigg(\dfrac{R_e} {\gamma}\bigg)^2}
\end{equation}
Now we use (23) and content of \textbf{Table 2.} to calculate the Intensity generated on Earth's surface.\textbf{Table 3.} shows theoretical results of intensity on Earth's surface. 
   \begin{table}[h]
      \caption[]{Total intensity for different Lorentz factor}
         \label{}
     $$ 
         \begin{array}{p{0.4\linewidth}l}
            \hline
            \noalign{\smallskip}
            Lorentz factor      &  Intensity(ergs/(s * cm^2))  \\
            \noalign{\smallskip}
            \hline
            \noalign{\smallskip}
            40 & 1.51326 * 10^{-8}   
            \\
            60    & 5.10729 * 10^{-8} 
            \\
            80    & 1.21062 * 10^{-7}       \\
            100   & 2.36450 * 10^{-7} 
            \\
            \noalign{\smallskip}
            \hline
         \end{array}
     $$ 
   \end{table}
  
  We are interested to see if the FAST telescope in China can detect this radiation. \textbf{Table 4.} shows effective power generated on FAST telescope using the effective area of detection $\pi R_{eff}^{2}$:
\begin{equation}
    P_{eff} = \pi \cdot R_{eff}^2 \cdot I
\end{equation}

      \begin{table}[h]
      \caption[]{Total effective power generated on FAST for different Lorentz factor}
         \label{}
     $$ 
         \begin{array}{p{0.4\linewidth}l}
            \hline
            \noalign{\smallskip}
            Lorentz factor      &  Effective \, Power(W))  \\
            \noalign{\smallskip}
            \hline
            \noalign{\smallskip}
            40 & 2.96599 * 10^{-6}   
            \\
            60    & 1.00103 * 10^{-5} 
            \\
            80    & 2.37282 * 10^{-5}       \\
            100   & 4.63442 * 10^{-5} 
            \\
            \noalign{\smallskip}
            \hline
         \end{array}
     $$ 
   \end{table}

Using the experimental data for Hubble telescope, which orbits the earth, we deduce that QLD related synchrotron radiation from Jupiter will reach earth's surface.\cite{NASA} ,\cite{Hubble}. Considering the fact that the system noise of the FAST telescope $T_{system} = 20 k$ and using the relation $P_{noise} = k_{boltzman} \cdot T_{system} \cdot Hz$  \cite{nan}. we calculate that $P_{noise} =2.7600 * 10^{-22}(watts)$ . Therefore based on \textbf{Table 4.} FAST should be able to detect given radiation.

\section{Conclusion}
    
    \, To conclude, we wanted to calculate frequency and the intensity range of  QLD related synchrotron radiation from ultralrelativistic electrons in Jupiter-Io system. 
    
    We started by modeling Jupiter's magnetic field and the concentration of the corotating electrons. As these particles move along the trajectory, they undergo the effect of forces.  We numerically calculated the results of these effects and showed how the equation for distribution function simplified. We also showed would would be range of the frequency of emitted photons. 
    
    Using the data for FAST telescope we determined that the radiation spectrum matched the detectable frequency range.\,Also the power generated by the radiation would be sufficient to surpass the power threshold, meaning that the FAST telescope will be able to detect the radiation.
    
    To summarise, we determined what would be the frequency range of the emission generated by ultrarelativistic electrons moving from Jupiter to Io torus if their nonzero pitch angles were sustained by QLD. Our results came in agreement with the empirical data.\,Finally, we numerically calculated the total power and the intensity on earth's surface.

\end{document}